\documentclass[aps,prl,twocolumn,superscriptaddress,
showpacs,amsmath,amssymb]{revtex4}

\usepackage{graphicx}
\usepackage{dcolumn}
\usepackage{bm}

\begin{document}

\title{Origin of the large thermoelectric power
in oxygen-variable $R$BaCo$_{2}$O$_{5+x}$ ($R$=Gd, Nd)}

\author{A. A. Taskin}
\affiliation{Central Research Institute of Electric Power
Industry, Komae, Tokyo 201-8511, Japan}

\author{A. N. Lavrov}
\affiliation{Institute of Inorganic Chemistry,
Novosibirsk 630090, Russia}

\author{Yoichi Ando}
\affiliation{Central Research Institute of Electric Power
Industry, Komae, Tokyo 201-8511, Japan}


\begin{abstract}

Thermoelectric properties of GdBaCo$_{2}$O$_{5+x}$ and
NdBaCo$_{2}$O$_{5+x}$ single crystals have been studied upon continuous
doping of CoO$_2$ planes with either electrons or holes. The thermoelectric
response and the resistivity behavior reveal a hopping character
of the transport in both compounds, providing the basis for understanding
the recently found remarkable divergence of the Seebeck coefficient at
$x=0.5$. The doping dependence of the thermoelectric power evinces that the
configurational entropy of charge carriers, enhanced by their spin and
orbital degeneracy, plays a key role in the origin of the large
thermoelectric response in these correlated oxides.

\end{abstract}


\maketitle

Materials with strong electron-electron interactions, called strongly
correlated electron systems, have recently attracted a great deal of
attention as a promising alternative to conventional semiconductors in the
field of thermoelectric power generation. The hope to find compounds
with superior thermoelectric properties among correlated materials is based
on the idea that in correlated systems, where the spin and orbital degrees
of freedom play an important role in transport properties, the
thermoelectric response can be enhanced by a large spin-orbital degeneracy
of charge carriers \cite{Chaikin, Koshibae, Kotliar}.

Na$_x$CoO$_2$ and several misfit-layered cobaltites with triangular-lattice
CoO$_2$ planes $\relbar$ materials demonstrating high thermoelectric
response and metallic behavior \cite{Terasaki} $\relbar$ have been
considered as possible examples of compounds where the spin-orbital
degeneracy plays a dominant role in enhancing the thermoelectric power
\cite{Koshibae, Terasaki}. However, a conventional Boltzmann-transport
approach can also explain the coexistence of a large Seebeck coefficient
and metallic conductivity in these compounds \cite{Singh}, blurring the
role and relevance of strong electron correlations.

Recently, layered cobaltites $R$BaCo$_{2}$O$_{5+x}$ (R is a rare-earth
element) with square-lattice CoO$_2$ planes came into focus because of
their remarkable transport and magnetic properties \cite{Martin, Respaud,
large_GBCO}. A variety of spin and orbital states are available in
$R$BaCo$_{2}$O$_{5+x}$ owing to its very rich phase diagram
\cite{large_GBCO}, where $x=0.5$ is the parent compound with all cobalt
ions in the 3+ valence state. The variability of oxygen content in
$R$BaCo$_{2}$O$_{5+x}$ allows one to dope continuously the CoO$_2$ planes
with either electrons (Co$^{2+}$ states) or holes (Co$^{4+}$ states) and to
measure a precise doping dependence of the Seebeck coefficient in one 
and the same crystal. Apart from this filling control, one can also employ a
bandwidth control of transport properties using rare-earth elements $R$
with different ionic radii [Fig. 1(a)] to clarify the role of
band structure parameters in determining the thermoelectric power in these
oxides. Thus, $R$BaCo$_{2}$O$_{5+x}$ can provide a suitable ground for
elucidating the relation between the spin-orbital degeneracy and
thermoelectric properties in correlated materials.

\begin{figure}[b]
\vspace{-6pt}
\includegraphics*[width=20pc]{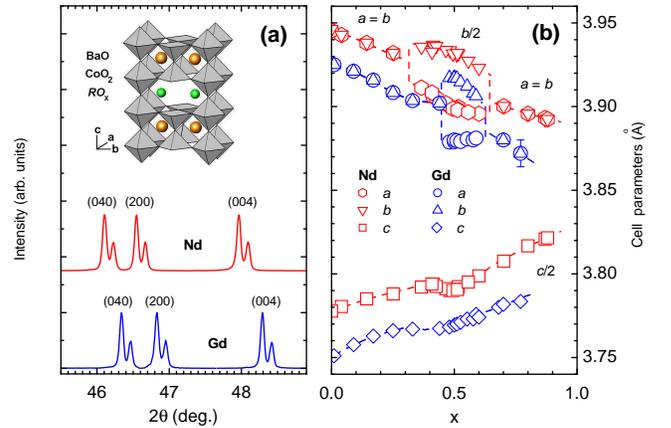}
\caption
{(Color online) (a) X-ray Bragg peaks (200), (040), and (004), measured in
GdBaCo$_{2}$O$_{5.5}$ and NdBaCo$_{2}$O$_{5.5}$ crystals at room
temperature (each peak has Cu $K\alpha_1$ and Cu $K\alpha_2$ contributions to
the diffraction pattern). Note that the unit cells are doubled along the
$b$ and $c$ axes. A sketch of the crystal structure of
$R$BaCo$_{2}$O$_{5.5}$ is shown in the upper part of the panel. (b) The
evolution of the room-temperature lattice parameters in
GdBaCo$_{2}$O$_{5+x}$ and NdBaCo$_{2}$O$_{5+x}$ under variation of the
oxygen content $x$.}
\label{fig1}
\end{figure}

Here we present a comparative study of the Seebeck coefficient
and the resistivity in GdBaCo$_{2}$O$_{5+x}$ (GBCO) and
NdBaCo$_{2}$O$_{5+x}$ (NBCO) single crystals over a wide range of electron
and hole doping. Despite the difference in the lattice parameters, both
compounds show virtually the same behavior of the Seebeck coefficient,
implying that the thermoelectric power in these compounds is governed by
correlated hopping of charge carriers rather than the band structure
parameters. An analysis of the doping dependence of the Seebeck
coefficient, showing a spectacular divergence at $x=0.5$, provides evidence
that the entropy contribution of charge carriers is the main source of the
large thermoelectric power in $R$BaCo$_{2}$O$_{5+x}$.

High-quality GdBaCo$_{2}$O$_{5+x}$ and NdBaCo$_{2}$O$_{5+x}$ single
crystals were grown using the floating-zone technique and their oxygen
content was modified by high-temperature annealing treatments with
subsequent quenching to room temperature \cite{large_GBCO}. Resistivity
measurements were carried out by a standard ac four-probe method. The
thermoelectric power was measured in a slowly oscillating thermal gradient
of $\sim 1$ K along the $ab$ plane. The contribution from the gold wires
($\sim 2$ $\mu$V/K) used as output leads was subtracted.

In parent $R$BaCo$_{2}$O$_{5.5}$ compounds, the crystal lattice consists of
equal numbers of CoO$_6$ octahedra and CoO$_5$ square pyramids
[schematically shown in the inset of Fig. 1(a)]. Upon changing the oxygen
content, some oxygen ions are inserted into or removed from the $R$O$_x$
planes, which changes the number of CoO$_6$ octahedra and CoO$_5$ pyramids
and also creates electrons or holes in CoO$_2$ planes. Figure 1(b) shows
the room-temperature lattice parameters in the entire available range of
oxygen concentrations. Regardless of the oxygen content, the unit cell
parameters in NBCO are found to be noticably larger than in GBCO, in
agreement with the larger ionic radius of Nd. This difference in the
unit-cell size implies the difference in Co$\relbar$O distances and/or
O$\relbar$Co$\relbar$O angles, which determine the one-electron bandwidth;
therefore, the substitution of the rare-earth element should affect the
thermoelectric properties, if they are governed by the band structure
parameters.

\begin{figure}[t]
\includegraphics*[width=20pc]{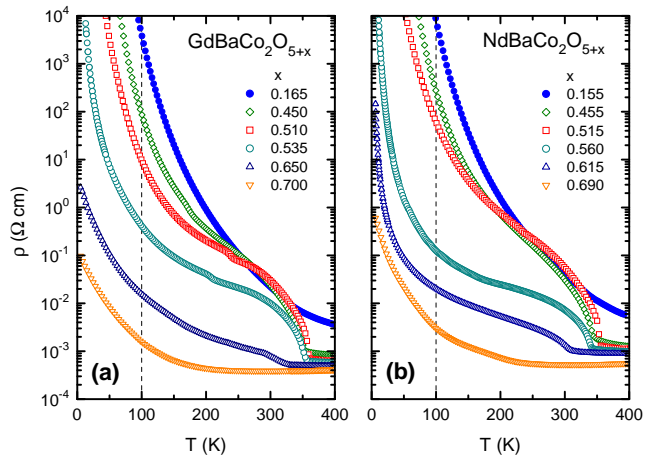}
\caption{(Color online) Temperature dependences of the in-plane
resistivity $\rho(T)$ of (a) GdBaCo$_{2}$O$_{5+x}$ and (b)
NdBaCo$_{2}$O$_{5+x}$ crystals.}
\label{fig2}
\end{figure}

Figure 2 shows the temperature dependences of the resistivity in GBCO and
NBCO for several oxygen concentrations. Both compounds demonstrate similar
$\rho (T)$ behavior, which strongly depends on the oxygen content. For
oxygen concentrations close to the parent composition $x=0.5$, $\rho (T)$
curves show a sharp metal-insulator (MI) transition upon cooling below
approximately the same temperature $T_{MI} \approx 360$ K, while the MI
transition seems to be smeared for $x$ away from $0.5$. Nevertheless, the
low-temperature resistivity exhibits a hopping character for the entire
range $0\leq x\leq 0.7$, which allows us to consider the charge transport
in terms of hopping of localized electrons (Co$^{2+}$) or holes
(Co$^{4+}$). In the insulating regime, the resistivity in both compounds
quickly decreases with hole doping, but remains unchanged or even increases
with electron doping. This doping asymmetry can be explained by the
different hopping probability of localized electrons (Co$^{2+}$) and holes
(Co$^{4+}$) moving in the background of Co$^{3+}$ ions because of the spin
blockade of the electron transport \cite{spin_blockade}.

A comparison of the temperature dependences of the Seebeck coefficient
$Q(T)$ in GBCO and NBCO for several oxygen concentrations reveals a
remarkable similarity of their thermoelectric properties (Fig. 3). Not only
the temperature dependences are similar, but also the absolute values of
$Q(T)$ are virtually the same in GBCO and NBCO. At high temperatures, both
compounds show a small, negative, and almost temperature-independent and
doping-independent Seebeck coefficient, which is quite natural for a
metallic state \cite{large_GBCO}. On the insulating side, on the other
hand, the temperature dependences of the Seebeck coefficient are rather
complicated and do not follow a simple $\sim 1/T$ law expected for
insulators. $Q(T)$ depends strongly on the oxygen content, being negative
for electrons and positive for holes, as shown in Fig. 4(a) for $T=100$ K.
Note that the absolute value of the Seebeck coefficient decreases rapidly
upon doping with electrons or holes, in contrast to the asymmetric doping
dependence of the conductivity, shown in Fig. 4(b) for the same
temperature. The most striking feature in the doping dependence of the
Seebeck coefficient $Q(x)$ [Fig. 4(a)] is its remarkable divergence at
$x=0.5$, where it reaches a large absolute value.

\begin{figure}[t]
\includegraphics*[width=20pc]{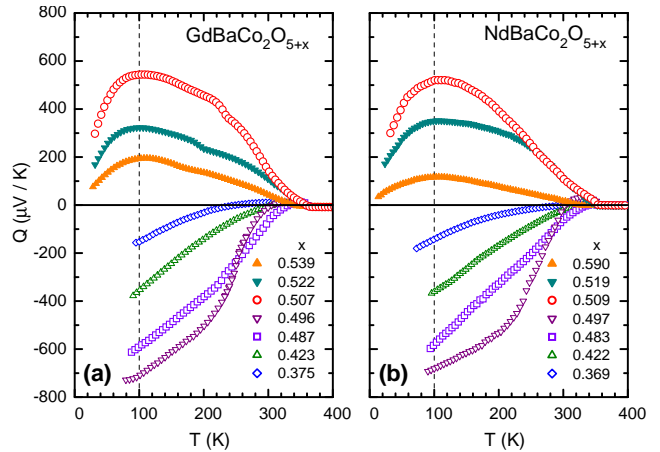}
\caption{(Color online) Temperature dependences of the Seebeck coefficient
$Q(T)$ of (a) GdBaCo$_{2}$O$_{5+x}$ and (b) NdBaCo$_{2}$O$_{5+x}$
crystals.}
\label{fig3}
\end{figure}

In order to clarify the origin of this large thermoelectric response, it is
useful to consider what one would expect for the doping behavior if
$R$BaCo$_{2}$O$_{5+x}$ were an ordinary band-gap material. Figure 4(c)
shows the plot of $Q$ versus the logarithm of the conductivity measured at
$T=100$ K -- the so called Jonker plot \cite{Jonker}. For band gap materials,
this plot is a curve with a universal shape [the pear-shaped curve, shown
in Fig. 4(c)], which can be scaled by intrinsic parameters related to the
energy band gap and scattering mechanism. In the extrinsic conduction
region, the curve $Q$ vs. ln($\sigma$) has a slope $\pm k_B/e$, where $k_B$
is the Boltzmann constant and $e$ is the absolute value of the electron
charge. Upon approaching the intrinsic-conduction region with decreasing
concentration of electrons or holes, this slope can only decrease. As can
be seen in Fig. 4(c), the actual slope of the Jonker plot for
$R$BaCo$_{2}$O$_{5+x}$ {\it increases} near the intrinsic conductivity
region, exceeding the ideal $k_B/e$ slope by several times. Moreover, the
decrease of $Q$ with electron doping does not correlate with the doping
dependence of the conductivity, which shows no increase because of the spin
blockade phenomenon \cite{spin_blockade}. Thus, both the temperature and
doping dependences of the Seebeck coefficient clearly deviate from the
behavior of conventional band gap materials.

Given the hopping character of the charge transport in
$R$BaCo$_{2}$O$_{5+x}$, one can reasonably assume that a large part of the
thermoelectric power comes from the entropy, which is carried by each
electron or hole along with a charge. To check this assumption, let us
examine the doping dependence of the Seebeck coefficient in GBCO and NBCO
crystals, $Q(x)$, measured at $T=100$ K [Fig. 4(a)]. Although the chosen
temperature is presumably too low to achieve the true high-temperature
limit, where the entropy contribution becomes the only contribution to the
thermoelectric power \cite{Koshibae}, it should be high enough to make
reasonable estimations. Note that for $R$BaCo$_{2}$O$_{5+x}$, which
undergoes a transition to a metallic state, the small and
doping-independent Seebeck coefficient at high temperatures is governed by
a different mechanism \cite{large_GBCO}.

The entropy contribution $\Delta Q$ to the thermoelectric power in
$R$BaCo$_{2}$O$_{5+x}$ should be proportional to the change in the entropy
$S$ of the electron system upon introducing $N$ charge carries at constant
internal energy $E$ and volume $V$ of the crystal \cite{Koshibae}, where
the entropy is determined by the total number of configurations $\Omega$,
i.e., all possible arrangements of introduced carriers in the crystal
lattice. Hence, $\Delta Q$ is expressed in terms of $\Omega$ as

\begin{equation}
\Delta Q=-\frac{1}{e}\left(  \frac{\partial S}{\partial N} \right)_{E,V}=
-\frac{k_B}{e} \frac{\partial ln\Omega}{\partial N} .
\end{equation}
Thus, the description of the doping dependence of the Seebeck coefficient
$Q(x)$ in our case can be reduced to a combinatorial problem.

\begin{figure}[t]
\includegraphics*[width=20.5pc]{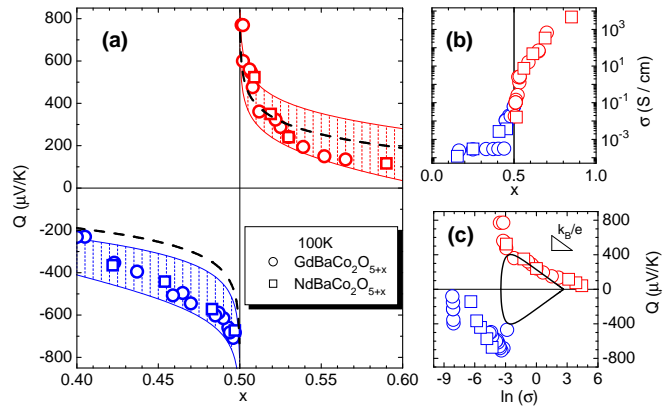}
\caption{(Color online) (a) The doping dependence of the Seebeck
coefficient $Q(x)$ of GdBaCo$_{2}$O$_{5+x}$ (circles) and
NdBaCo$_{2}$O$_{5+x}$ (squares) crystals at $T=100$ K. The hatched area
represents the entire available range of the entropy contribution to the
thermopower $\Delta Q_n(x)$ and $\Delta Q_p(x)$. (b) The doping dependence
of the conductivity $\sigma (x)$ of GdBaCo$_{2}$O$_{5+x}$ (circles) and
NdBaCo$_{2}$O$_{5+x}$ (squares) crystals at $T=100$ K. (c) Jonker plot for
$R$BaCo$_{2}$O$_{5+x}$ (see text).}
\label{fig4}
\end{figure}

In the simplest case of nondegenerate carriers, the above approach gives
the celebrated Heikes formula \cite{Chaikin}
\begin{equation}
\Delta Q_n= -\frac{k_B}{e} \;ln\left[ \frac{\frac{1}{2}+x}{\frac{1}{2}-x}\right], \;
\Delta Q_p= \frac{k_B}{e} \;ln\left[ \frac{\frac{3}{2}-x}{x-\frac{1}{2}}\right] .
\end{equation}
Both electron $\Delta Q_n$ and hole $\Delta Q_p$ contributions are shown in
Fig. 4(a) by the dashed lines. A comparison with experimental data shows
that this simple model, albeit it gives a rough idea of why the Seebeck
coefficient can diverge at $x=0.5$, is not sufficient for a quantitative
description of the doping dependence $Q(x)$ in $R$BaCo$_{2}$O$_{5+x}$,
especially in the case of electron doping.

As has been recently emphasized by Koshibae and Maekawa \cite {Koshibae},
the spin-orbital degeneracy of charge carriers plays an important role in
systems with correlated hopping transport. In $R$BaCo$_{2}$O$_{5+x}$, there
are two kinds of lattice sites for cobalt ions that differ in the local
oxygen environment [see the inset of Fig. 1(a)], namely, $x$ octahedra and
$(1-x)$ pyramids, which provide different spin-orbital states for
introduced electrons (Co$^{2+}$), holes (Co$^{4+}$), or host Co$^{3+}$
ions. For electron doping, if one knows the distribution of introduced
electrons (Co$^{2+}$) between octahedral and pyramidal positions, the total
number of configurations $\Omega$ is determined by

\begin{equation}
\Omega = g_{2o}^{N_{2o}}g_{2p}^{N_{2p}}g_{3o}^{N_{3o}}g_{3p}^{N_{3p}}
\cdot \frac{N_L!}{N_{2o}!\;N_{2p}!\;N_{3o}!\;N_{3p}!} ,
\end{equation}
where $N_L$ is the total number of cobalt sites; $N_{2o}$ and $N_{2p}$
($N_{3o}$ and $N_{3p}$) are the numbers of Co$^{2+}$ (Co$^{3+}$) ions in
octahedral and pyramidal positions, respectively; $g_{2o}$, $g_{2p}$,
$g_{3o}$, $g_{3p}$ are the degeneracies of these states. An analogous
expression can be written for hole doping as well.

At any given doping level in the range of $0 \leq x \leq \frac{1}{2}$, the
equilibrium concentrations of Co$^{2+}$ and Co$^{3+}$ ions in different
positions are determined by the only parameter -- a probability $p$ to find
a Co$^{2+}$ ion in an octahedral position. For $x$ close to $x=0.5$, where
the number of doped carries is not large, this probability is determined by
the law of mass action ($N_{2p} \cdot N_{3o})/(N_{2o} \cdot N_{3p})=K(T)$
\cite{mass_act}, which couples the equilibrium concentrations of Co$^{2+}$
and Co$^{3+}$ ions that are competing for octahedral and pyramidal
positions. This equation can be analytically solved for $p$, yielding
\begin{equation}
p=\frac{\sqrt[]{(\frac{K+1}{2})^2 +4(K-1) \cdot (\frac{1}{2}-x)
\cdot x}-\frac{K+1}{2}}{2 (K-1) \cdot (\frac{1}{2}-x)},
\end{equation}
where $K(T)$ is the equilibrium constant, which depends only on the
temperature and intrinsic characteristics of the system, such as the
difference in energies of Co$^{2+}$ (and Co$^{3+}$) ions in octahedral and
pyramidal positions.

Using Eqs. (1), (3), and (4), the analytical solution for the entropy
contribution of doped electrons to the thermoelectric power can be written
as follows:
\begin{equation}
\Delta Q_n=-\frac{k_B}{e}\;ln\left[ \frac{g_{2o}^{m}\;\;\;\;
g_{2p}^{1-m}}{g_{3o}^{1+m}\;g_{3p}^{-m}}
\cdot \frac{N_{3o}^{1+m}\;N_{3p}^{-m}}{N_{2o}^{m}\;\;\;\;
N_{2p}^{1-m}} \right],
\end{equation}
\\
where $m \equiv p+ (x-\frac{1}{2}\;) \cdot \frac{\partial p}{\partial x}$.
Upon electron doping, $m$ can change its value in the range from -1 to +1,
depending on the intrinsic characteristics of the system. The entropy
contribution of doped holes to the thermoelectric power in the range
$\frac{1}{2} \leq x \leq 1$ can be obtained in the same way.

As follows from Eq. (5), there are only two factors that affect the doping
dependence of the Seebeck coefficient: The equilibrium constant $K$, which
determines the occupation of octahedral (pyramidal) positions by doped
electrons, and degeneracies of states $g_i$, which turn out to be the main
source of the thermoelectric-power enhancement in $R$BaCo$_{2}$O$_{5+x}$
near $x=0.5$. Note that the entropy contribution to the thermoelectric
power is expected to be insensitive to the unit-cell size, which gives a
natural account for the similarity of the Seebeck coefficient $Q(T)$ in
GBCO and NBCO.

It is well known that owing to a comparable strength of the exchange
interaction and the crystal field splitting, Co$^{3+}$ ions can adopt
different spin states, depending on the oxygen environment or even on
temperature \cite{LaCoO3}. It is believed that in the parent
$R$BaCo$_{2}$O$_{5.5}$ at low temperature, the Co$^{3+}$ ions adopt the
low-spin (LS) state in octahedra and the intermediate-spin (IS) state in
pyramids \cite{Respaud, large_GBCO}. For Co$^{2+}$ the crystal field is
weaker than for Co$^{3+}$ and thus Co$^{2+}$ adopts the high-spin (HS)
state, while for Co$^{4+}$ the crystal field is stronger, favoring the LS
state. Any state with nonzero spin $S$ has the degeneracy $2S+1$. In
addition to this spin degeneracy, there can be orbital degeneracy as well.
For both HS-Co$^{2+}$ and LS-Co$^{4+}$ the orbital state is threefold
degenerate because of a hole residing in the $t_{2g}$ orbitals. In general,
the orbital degeneracy can be lifted by lowering the crystal symmetry. For
a rough estimation, we assume that the orbital degeneracy in pyramidal
positions is lifted, which gives the following spin-orbital degenaracies:
$g_{2o}=12$, $g_{2p}=4$, $g_{3o}=1$, $g_{3p}=3$, $g_{4o}=6$, and
$g_{4p}=2$.

The hatched area in Fig. 4(a) shows the entire available range of $\Delta
Q_n(x)$ and $\Delta Q_p(x)$ for $R$BaCo$_{2}$O$_{5+x}$, which is obtained
by changing the equilibrium constant $K$ from one extreme to another (where
$p$ changes from 0 to 1) and keeping the spin-orbital degeneracies
unchanged. As can be seen in Fig. 4(a), all experimental values of the
Seebeck coefficient lie within this area. This gives confidence that the
entropy contribution of charge carriers, which includes their spin-orbital
degeneracy, can account for the doping dependence of the Seebeck
coefficient and its remarkable divergence at $x=0.5$. Also, this model
naturally accounts for the somewhat enhanced thermoelectric response in
electron-doped crystals in comparison with the hole-doped ones: The larger
Seebeck coefficient for electrons can be explained by the larger degeneracy
of HS-Co$^{2+}$ in comparison with LS-Co$^{4+}$.

In conclusion, the present study of thermoelectric and transport properties
in $R$BaCo$_{2}$O$_{5+x}$ gives a solid experimental support to the idea
that strong electron correlations and spin-orbital degeneracy can bring
about a large thermoelectric power in transition-metal oxides.



\begin{thebibliography}{99}

\bibitem{Chaikin} P.M. Chaikin and G. Beni,
Phys. Rev. B {\bf 13}, 647 (1976).

\bibitem{Koshibae} W. Koshibae, K. Tsutsui, and S. Maekawa,
Phys. Rev. B {\bf 62}, 6869 (2000);
W. Koshibae and S. Maekawa,
Phys. Rev. Lett. {\bf 87}, 236603 (2001).

\bibitem{Kotliar} V. S. Oudovenko and G. Kotliar,
Phys. Rev. B {\bf 65}, 075102 (2002).

\bibitem{Terasaki} I. Terasaki, Y. Sasago, and K. Uchinokura,
Phys. Rev. B {\bf 56}, R12685 (1997);
Y. Ando, N. Miyamoto, K. Segawa, T. Kawata and I. Terasaki, 
Phys. Rev. B {\bf 60}, 10580 (1999);
Y. Wang, N. S. Rogado, R. J. Cava, and N. P. Ong,
Nature {\bf 423}, 425 (2003).

\bibitem{Singh} D. J. Singh, Phys. Rev. B {\bf 61}, 13397 (2000);
T. Takeuchi, T. Kondo, T. Takami, H. Takahashi, H. Ikuta, U. Mizutani, 
K. Soda, R. Funahashi, M. Shikano, M. Mikami, S. Tsuda, T. Yokoya, 
S. Shin, and T. Muro, 
Phys. Rev. B {\bf 69}, 125410 (2004).

\bibitem{Martin} C. Martin, A. Maignan, D. Pelloquin, N. Nguyen,
and B. Raveau, Appl. Phys. Lett. {\bf 71}, 1421 (1997).

\bibitem{Respaud} M. Respaud, C. Frontera, J. L. Garc\'{i}a-Mu\~{n}oz,
M. A. G. Aranda, B. Raquet, J. M. Broto, H. Rakoto, M. Goiran, A. Llobet, 
and J. Rodr\'{i}guez-Carvajal, Phys. Rev. B {\bf 64}, 214401 (2001).

\bibitem{large_GBCO} A. A. Taskin, A. N. Lavrov, and Y. Ando,
Phys. Rev. B {\bf 71}, 134414 (2005).

\bibitem{spin_blockade} A. Maignan, V. Caignaert, B. Raveau, D. Khomskii,
and G. Sawatzky, Phys. Rev. Lett. {\bf 93}, 026401 (2004);
A. A. Taskin and Y. Ando,
Phys. Rev. Lett. {\bf 95}, 176603 (2005).

\bibitem{Jonker} G.H. Jonker, Philips Res. Reps. {\bf 23}, 131 (1968);
G.M. Choi, H.L. Tuller, and D. Goldschmidt,
Phys. Rev. B {\bf 34}, 6972 (1986).

\bibitem{mass_act} F.A. Kr\"oger,
{\it The Chemistry of Imperfect Crystals} (North-Holland, Amsterdam, 1974).

\bibitem{LaCoO3} P.M. Raccah and J.B. Goodenough,
Phys. Rev. {\bf 155}, 932 (1967).

\end{thebibliography}
\end{document}